\documentclass[showpacs, pra,twocolumn,preprintnumbers ,amsmath, amssymb, superscriptaddress, aps]{revtex4-2}
\usepackage{anyfontsize}
\usepackage{helvet}    
\usepackage{color}
\usepackage{amsmath,amssymb}
\usepackage{pifont}
\usepackage{amssymb}  
\usepackage{bbold}
\usepackage{float}
\usepackage{subfloat}
\usepackage[caption=false]{subfig}
\usepackage{tikz}
\usepackage{makecell}
\usepackage{subfig}
\usepackage{pifont}   
\usepackage{graphicx} 
\graphicspath{{Figuress/}}
\usepackage{dcolumn}  
\usepackage{bm}       
\usepackage{multirow} 
\usepackage{placeins}
\usepackage[colorlinks]{hyperref}
\usepackage{mathtools}
\usepackage{appendix}

\newcommand{\bb} {\color{blue}}

\def \be{\begin{align}}
	\def \ee{\end{align}}
\def \bea{\begin{eqnarray}}
	\def \eea{\end{eqnarray}}

\begin{document}
	\renewcommand{\thesection}{\arabic{section}}
	\renewcommand{\thesubsection}{\arabic{section}.\arabic{subsection}}
	\renewcommand{\thefigure}{\arabic{figure}}
	\baselineskip=0.9cm

	\title{Comparative analysis of real experiments and digital (ICT) simulations regarding their impact on student learning.}
	
	\author{Rachid El Aitouni}
		\email{elaitouni.r@ucd.ac.ma}
	\affiliation{Laboratory of Theoretical Physics, Faculty of Sciences, Choua\"ib Doukkali University, PO Box 20, 24000 El Jadida, Morocco}
	\author{Ahmed Bouhlal}
	\affiliation{Laboratory of Theoretical Physics, Faculty of Sciences, Choua\"ib Doukkali University, PO Box 20, 24000 El Jadida, Morocco}
	
	\begin{abstract}
This study, conducted among more than 250 physics and chemistry teachers in Morocco, analyzes the impact of experimentation on student learning and attention in middle and high school. The results show that the majority of teachers favor digital simulations, except for simple experiments such as electrical circuits. This choice is linked to material constraints, class size, and safety requirements. 
Simulations are perceived as practical and flexible, allowing experiments to be repeated or slowed down to facilitate understanding. However, teachers emphasize the need for specific ICT training in order to better integrate these tools into their practices. The most effective strategy identified is based on a hybrid approach: using simulations to explain abstract phenomena and real experiments to develop experimental skills, methodological rigor, and critical thinking. This complementary approach appears to be a promising solution for enriching science education and overcoming the constraints encountered in schools.\\

Key Words: Experience, ICT, simulation, middle school, high school
			\end{abstract}
	
	\maketitle
	
	\section{Introduction}	\label{Intro}
Physics and chemistry teaching based on activities evolves progressively with increasing difficulty in concepts and calculations. Most of the chapters in the first year of middle school are based on simple analysis and the students’ environment. Many of these concepts have already been introduced in elementary school, but they are revisited with more detail in the first year. In the second-year program, the level of difficulty increases with the introduction of new concepts such as atoms and chemical reactions. In the third year, additional notions such as ions are introduced, aligning the level of complexity with that of high school. For the final year of high school, the problems become more complex than at other levels, requiring advanced reasoning and analytical skills.  

Teaching methods are adapted to the students’ level of understanding and cognitive development. At the middle school cycle, the essential teaching method is \textbf{investigation} ({\bb Lougman,et al., 2023; De Hosson et al,. 2016}), since students at this stage are capable of grouping ideas and reconstructing solutions to new situations in a simple and clear way. This approach is based on understanding the situation, proposing a research question, and then forming hypotheses that are verified either experimentally, through observation, or by consulting references. For first-year students, reformulating the research question remains limited, requiring teacher support. However, at higher levels of secondary school, the research process becomes more advanced and problem-solving emerges as the dominant method. This technique requires concentration and the ability to recombine information to find solutions, with hypotheses most often verified through experimentation.  

Experimentation is widely recognized as one of the \textbf{fundamental pillars of science education} (\textcolor{blue}{Millar, 2004; Hofstein \& Lunetta, 2004; Hodson, 1990}). It is considered the most effective way to bring abstract concepts closer to learners’ minds. Direct experience through real experiments consolidates scientific knowledge, develops observation and analysis skills, and promotes initiative and inquiry. However, several studies have highlighted the \textbf{constraints of laboratory practice}, such as lack of equipment, overcrowded classrooms, and insufficient training (\textcolor{blue}{Taoufik et al., 2016; Mazouze et al., 2015; Caillods et al., 1998}). These obstacles often limit the feasibility of real experimentation in schools.  

With the rapid development of information and communication technologies, \textbf{digital simulations} have emerged as complementary educational tools. They allow learners to interact with scientific phenomena in a safe, flexible, and repeatable environment, offering clear visualization of abstract or invisible processes (\textcolor{blue}{Niedderer et al., 2002; Hucke \& Fischer, 2002}). Simulations are particularly valued for their ability to reduce risks, save time, and provide accessibility when material resources are lacking. Nevertheless, most scholars agree that simulations cannot fully replace real experiments, but rather serve as a \textbf{complementary approach} (\textcolor{blue}{Hassouni et al., 2014; Kane, 2011}).  

The combination of real experimentation and simulation has thus become a central topic in educational research. It directly impacts both the professional performance of teachers and the academic achievement of learners (\textcolor{blue}{Hofstein \& Lunetta, 2004; Abid et al., 2022}). This research seeks to study the relationship between these two types of experimentation by analyzing classroom practices and monitoring their impact on teaching quality and student motivation. It also aims to highlight the advantages and challenges associated with each method and to propose ways to integrate them in order to achieve a balance between realism and technological effectiveness, thereby enhancing the overall quality of teaching and learning in physics and chemistry.  The survey results show that real-life experiments generate more interest among learners than simulations. Although simulations are simpler and quicker to implement, and allow the pace of experiments to be slowed down or sped up, understanding is more effective through direct experimentation. However, several constraints limit the implementation of real-life experiments, such as a lack of equipment in laboratories, overloaded curricula, and large class sizes. Teachers also express a need for continuing education. Among the solutions proposed to increase the use of real experiments, which are essential for teaching and understanding, are equipping laboratories, organizing practical work in groups, and adding additional sessions. Ultimately, the two approaches—real experimentation and simulation using ICT—appear to be complementary, and combining them significantly improves learners' comprehension rates.

This paper is organized as follows: in the section 2, we present the working strategy and explain the rationale behind the choice of questionnaire items. In the section 3, we analyze the results, linking them to the identified constraints and proposing applicable solutions for conducting experiments, whether real or simulated. Finally, the section 4, provides a general conclusion to the study.

\section{Methodology}
In our study, which aims to evaluate the preference for traditional experimentation (real experience) and simulation, i.e., the use of Information and Communication Technologies for Education (ICT) and their impact on the quality of teaching, based on responses from more than 264 physics and chemistry teachers. To administer the questionnaire to participants, we used an online questionnaire. The use of questionnaires is a widely recognized method in educational and social science research. This methodological tool has several advantages: it allows for the collection of reliable data on the perceptions, attitudes, and behaviors of a large number of participants, while remaining economical, quick, and easy to administer. In addition, it standardizes responses, which facilitates comparison and statistical analysis. The questionnaire consisted of 18 questions, divided into two parts. First, we asked about the teaching cycle, gender, and age of the learners, followed by questions focused on the choice of experimental method, either conducting real experiments with learners in the institution's laboratory or using digital experiments through simulation, i.e., the integration of ICT in teaching practices, as well as the effectiveness of each type of experiment. 

This approach not only highlights general trends in the choice of experimentation methods, but also identifies the specific constraints faced by teachers (lack of equipment, large class sizes, limited session time, etc.) and helps to understand how these factors influence their teaching choices and the quality of teaching using each method. In the following section, we analyzed the results obtained from the questionnaires. Generally, the participation of a larger number of teachers provides a very clear picture of the quality of learning and the preferences for experiments carried out in the classroom, the level of understanding of the learners, and the satisfaction of the teachers.
\section{Results Analysis}
\subsection{Profile of Respondents}
\begin{figure}
\subfloat[]{\includegraphics[scale=0.505]{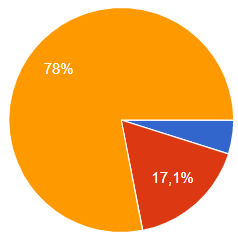}\label{years}}
\subfloat[]{\includegraphics[scale=0.5]{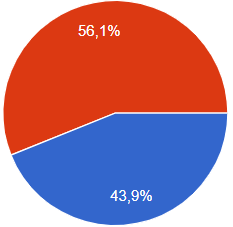}\label{school}}
\subfloat[]{\includegraphics[scale=0.5]{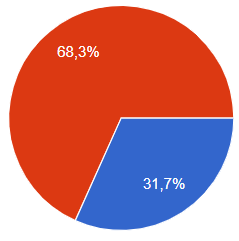}\label{sexe}}
\caption{Statistics of participating teachers. (a): Less than 5years (blue), between 5 and 10 years (red), more the 10 year (orange). (b): Midlle school (blue), high school (red). (c): Female (blue), male (red)}\label{stat}
\end{figure}

According to the results obtained, it appears that the majority of participants have more than ten years of professional experience in teaching (Fig. \ref{years}). This seniority lends significant credibility to the responses collected, as these teachers have had time to observe, analyze, and gain an in-depth understanding of the difficulties encountered in the classroom. Their perspective is therefore marked by a pedagogical maturity that considerably enriches the value of the data collected. Their long experience has enabled them to develop a detailed understanding of the constraints associated with teaching science, whether in terms of managing class sizes, using material resources, or adapting teaching methods to the diverse profiles of their students. In this sense, their testimonies are not limited to specific impressions, but reflect genuine expertise built up over many years. The study sample includes middle school and high school teachers, covering two essential levels of schooling. The presence of these two categories of teachers provides a more complete picture of teaching practices, taking into account the specific characteristics of each cycle (Fig. \ref{school}). However, there is a slight predominance of high school teachers, which reflects the structure of the sample and may be related to the availability or particular interest of this category in the subject under study.
Gender diversity was also respected, as both male and female teachers contributed to the survey (Figure \ref{sexe}). This diversity is important because it ensures a more balanced representation and allows for a variety of perspectives to be taken into account. The differences in background, sensibilities, and sometimes priorities between the two genders enrich the analysis and avoid a biased view. Finally, several teachers reported contrasting material conditions (Figure \ref{obstacles}): some have partially equipped laboratories, while others mentioned a total lack of functional laboratories. This observation is particularly significant, as it directly affects the possibility of conducting real experiments in the classroom. The absence or insufficiency of scientific equipment is a major constraint that influences teaching choices and may lead teachers to make greater use of digital simulations. Thus, the profile of the respondents, characterized by solid experience, a diversity of teaching levels, and gender diversity, provides a rich and credible basis for analysis. The material conditions reported add an essential dimension to the understanding of practices, showing that pedagogical choices depend not only on teachers' preferences, but also on available resources and institutional constraints.

\subsection{Challenges between real experiments and ICT-based simulations}
	\begin{figure}
		\subfloat[]{\includegraphics[scale=0.505]{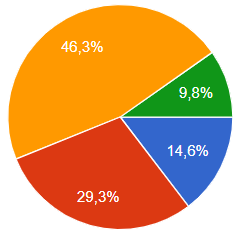}\label{reel}}
		\subfloat[]{\includegraphics[scale=0.5]{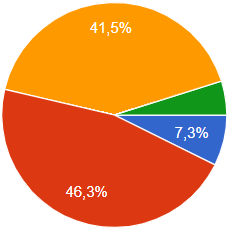}\label{Tice}}
		\subfloat[]{\includegraphics[scale=0.5]{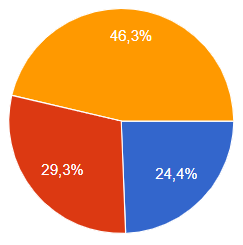}\label{labo}}\\
			\subfloat[]{\includegraphics[scale=0.505]{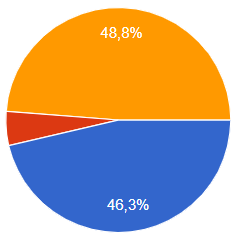}\label{comp}}
		\subfloat[]{\includegraphics[scale=0.5]{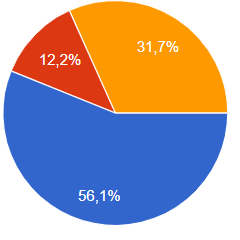}\label{motiv}}
		\subfloat[]{\includegraphics[scale=0.5]{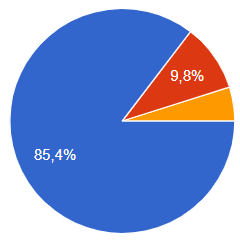}\label{obstacles}}
		\caption{Using real experiments or simulations. (a): Real experiments. (b): Digital simulations (ICTE).  Very often (blue), often (red), rarely (orange), never (green).
(c): Laboratory equipment. To equip (blue), unequipped (red), partially equipped (orange). (d): the most effective method for promoting students' understanding. Real experiments (blue), Digital simulations (red), both (orange). (e): The teaching method that motivates students, Real experiments (blue), Digital simulations (red), both (orange). (f): The constraints of real experiments, lack of equipment (blue), lack of time (red), safety constraints (orange).}\label{stat}
	\end{figure}	
According to the results of our survey, it appears that real-world experiences are generally used more often, as illustrated in Figure \ref{reel}. However, approximately $10\%$ of teachers report never implementing this type of activity. The main reason given is a lack of equipment, as shown in Figure \ref{obstacles}. In fact, $85\%$ of respondents say that their laboratories are completely lacking in equipment, which is a major obstacle to carrying out practical work. Regarding digital experiments based on ICT, $41.5\%$ of teachers report using them. The majority of those who use this type of experiment are secondary school teachers, as at these levels the experiments become more complex and require greater precision. For this reason, teachers prefer to use digital simulations, even when equipment is available, as shown in Figure \ref{Tice}. In middle school, the situation is different: the simplicity of the experiments offered makes it easier to carry out real experiments, even if digital alternatives exist. Teachers therefore favor hands-on activities, as they remain accessible and suited to the educational objectives of this level.

Figure \ref{comp} highlights that real-life experiments are highly appealing to learners. Indeed, $56\%$ of survey participants confirm that this type of activity generates particular interest and promotes student engagement. This preference can be explained by the concrete and lively nature of scientific experiments, which allow learners to move from theory to practice, verify the phenomena studied for themselves, and develop active curiosity. Real-life experiments thus offer an immersive dimension that stimulates motivation and reinforces understanding of concepts. However, despite this undeniable appeal, several major constraints limit the regular implementation of these activities in the classroom. The first difficulty lies in the lack of equipment available in school laboratories. Many teachers report that the necessary equipment is either non-existent or insufficient, making it impossible to carry out certain experiments. This lack of equipment is a structural obstacle that considerably reduces teaching opportunities.

Added to this constraint is the heavy workload of school curricula. Teachers must cover a large amount of content in a limited amount of time, leaving little room for organizing experimental sessions. However, real experiments require careful preparation, time to set up, carry out, and analyze the results. This time requirement often conflicts with the pace imposed by the curriculum, forcing teachers to favor faster methods, such as demonstrations or digital simulations. Finally, large class sizes are another major difficulty. Supervising a large number of students during an experimental activity requires rigorous organization and appropriate material conditions. Safety management, monitoring of manipulations, and individualized support become particularly complex when classes are overcrowded. This situation leads some teachers to limit the use of real experiments, despite their recognized educational value.

Thus, although real experiments are perceived as more attractive and educational by a majority of learners, their implementation remains hampered by material, time, and organizational constraints. These obstacles explain why, in many contexts, teachers turn to more accessible alternatives, such as digital simulations or group demonstrations. Nevertheless, the interest shown by students highlights the importance of finding solutions to strengthen the place of real-life experiments in physics and chemistry teaching, as they are an essential lever for motivation and the development of practical skills.

\subsection{Contexts in which simulation or real experiments are used}
	\begin{figure}
\subfloat[]{\includegraphics[scale=0.45]{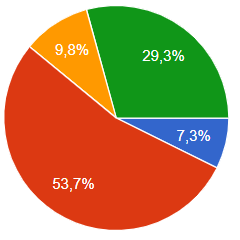}\label{simplereel}}
\subfloat[]{\includegraphics[scale=0.45]{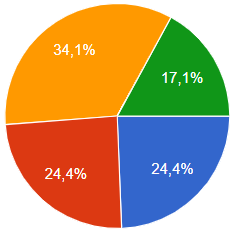}\label{simpleTice}}
	\subfloat[]{\includegraphics[scale=0.45]{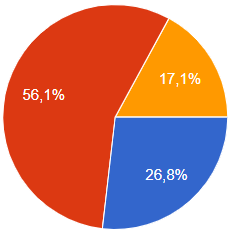}\label{facile}}
		\subfloat[]{\includegraphics[scale=0.45]{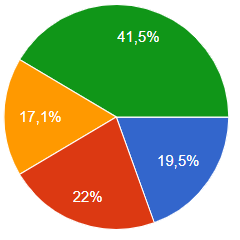}\label{avantagede TICE}}
	\caption{(a): The part of the course best suited to real experiments. (b): The part of the course best suited to digital simulations. Mechanics (blue), electricity (red), optics (orange), Chemistry (green). (c): The simplest method. Real experiments (blue), Digital simulations (red), both (orange). (d): Advantages of digital simulations. Accessibility (blue), speed (red), repeatability (orange), clear visualization of invisible phenomena (green). }
\end{figure}	
Figure illustrates participants' responses regarding which parts of the course are considered most suitable for conducting real experiments (Fig{\ref{simplereel}) and which are more suited to digital simulations (Fig{\ref{simpleTice}). Analysis of the results highlights a clear distinction between scientific disciplines in terms of their practical feasibility in the laboratory and their potential for modeling using digital tools.

The teachers surveyed consider the section on electricity to be the easiest to implement in school laboratories (about $34\%$). This ease can be explained by several factors. On the one hand, the components needed to carry out electrical experiments—such as batteries, wires, light bulbs, switches, and resistors—are common items, available in most schools and well known to the teaching community. On the other hand, electrical experiments are generally simple to perform and do not require overly restrictive safety conditions. Teachers can therefore offer students hands-on activities, such as building electrical circuits, measuring current and voltage, or studying the fundamental laws of electricity (Ohm's law, node law, mesh law). These accessible and reproducible experiments allow learners to develop practical skills while consolidating their understanding of theoretical concepts. However, when it comes to the part of the course devoted to optics, teachers believe that digital simulations are the most suitable tool. Optics involves phenomena that are more complex to reproduce in a school laboratory. Simulation allows the trajectory of light from the source to the receiver to be visualized accurately, which is often difficult to achieve with limited equipment. Teachers can thus show students a variety of physical phenomena, such as eclipses, the formation of shadows, the reflection and refraction of light, and the formation of images by mirrors and lenses. Although these experiments are fundamental to understanding the laws of optics, they are often difficult to perform in a school laboratory due to the need for specific equipment (high-quality lenses, suitable light sources, and precise measuring devices). Digital simulations offer an effective alternative, overcoming these difficulties while providing a clear and accessible visualization of the phenomena being studied.
The majority of teachers surveyed say that simulations are easier to implement than real experiments, as shown in Fig. \ref{facile}. This preference can be explained by the flexibility offered by digital tools: they allow an experiment to be viewed from multiple angles, repeated as many times as necessary, and adapted to suit educational needs. Simulations also offer the possibility of slowing down or speeding up the course of a phenomenon, which facilitates understanding of key stages. For example, in the case of optics, it is possible to represent the propagation of light in a gradual manner, to show the effect of a change in the angle of incidence, or to modify the nature of the medium being traversed. These features enrich teaching and enable students to better understand sometimes abstract concepts. Fig. \ref{avantagede TICE} confirms this trend by showing that participants particularly appreciate the ability to repeat a digital experiment an unlimited number of times. This repeatability is a major advantage over real experiments, which require preparation time, equipment verification, and sometimes cumbersome logistical organization. Digital simulations, on the other hand, can be restarted instantly, which encourages learning by trial and error and allows students to consolidate their knowledge by observing the same phenomenon several times. This interactive dimension helps to reinforce learners' motivation, as they feel freer to explore and experiment without fear of making irreversible mistakes. It should be noted, however, that this preference for simulations does not mean that real experiments are rejected. Teachers recognize the irreplaceable value of hands-on experimentation, which allows students to develop practical skills and confront the uncertainties of the scientific process. Nevertheless, in a context marked by material and organizational constraints, simulations appear to be a pragmatic and effective solution for maintaining the quality of science education. They make it possible to compensate for the shortcomings of school laboratories while offering students a rich and interactive educational experience.

Ultimately, analysis of the results shows that electricity is perceived as the most accessible discipline for real experiments, while optics is considered more suitable for digital simulations. This distinction reflects both the availability of equipment and the complexity of the phenomena studied. It also highlights the growing importance of digital tools in science education, not as exclusive substitutes for real experiments, but as indispensable complements to enrich teaching practices and respond to constraints in the field.
\section{Conclusion}
Reflection on the use of real-life experiments and digital simulations in science education highlights a constant tension between material and organizational constraints on the one hand, and educational ambitions on the other. The teachers interviewed emphasize that, despite the undeniable appeal of real-life experiments for students, their implementation faces major obstacles. Lack of equipment, overloaded curricula, and large class sizes are structural barriers that limit the frequency and quality of experimental activities. These difficulties are not anecdotal: they reflect a reality experienced in many schools where laboratories are inadequately equipped or even non-existent, and where teachers have to cope with working conditions that do not always allow students the opportunity to manipulate and experiment. In this context, digital simulations appear to be a pragmatic solution, capable of maintaining educational continuity and guaranteeing a minimum level of exposure to scientific phenomena, even in the most disadvantaged environments.

The advantages of simulations are numerous and widely recognized by teachers. They offer the possibility of repeating an experiment as many times as necessary, which promotes the consolidation of learning. They also make it possible to visualize abstract or invisible phenomena, making concepts accessible that would otherwise be difficult to grasp. Their speed and accessibility make them tools that are well suited to the time constraints and realities of large classes. Finally, they significantly reduce safety risks by preventing students from being exposed to dangerous manipulations or sensitive chemicals. These advantages explain why digital simulations are increasingly being integrated into teaching practices and why they are seen as an indispensable tool in contexts where material resources are limited. They are not only presented as a substitute alternative, but as a complementary modality that enriches teaching and opens up new didactic perspectives.

However, the question of whether simulations can completely replace real experiments is the subject of nuanced debate. The majority of teachers surveyed believe that simulations, however effective they may be, cannot entirely replace real experiments. The reason is simple: they do not allow students to develop the practical skills, methodological rigor, and critical thinking that come from direct confrontation with experimental reality. Manipulation, observation of possible errors, and management of uncertainties are essential steps in students' scientific training, and they cannot be authentically reproduced by a digital simulation. However, a minority of teachers recognize that simulations can play a partial substitute role, particularly in situations where safety or precision is at stake. But the consensus remains clear: real-life experiments remain irreplaceable for training students to think scientifically and develop solid practical skills.
This observation leads to broader reflection on the need to support teachers in integrating digital technologies. A significant proportion of them express the need for continuing education in order to better understand how to use ICT effectively and appropriately. This demand reflects a desire to adapt to technological developments and an openness to pedagogical innovation. Teachers do not reject digital tools, but seek support in integrating them into their practices without losing sight of the fundamental objectives of science education. Continuing education thus appears to be an essential lever for building teachers' confidence, improving their mastery of digital tools, and promoting a balanced use of simulations and real-life experiments. It is a prerequisite for teachers to be able to fully exploit the potential of ICT while preserving the irreplaceable value of hands-on activities.

With this in mind, several teachers propose an optimal combination that integrates simulations and real-life experiments into a coherent teaching sequence. This hybrid approach consists of starting with a simulation to introduce concepts, visualize abstract phenomena, and mentally prepare students. It is followed by a real-life experiment, which allows learners to manipulate, observe, and confront the constraints of scientific practice. Finally, it returns to the simulation to model, analyze, compare results, and generalize concepts. This strategy is perceived as the most effective because it maximizes both conceptual understanding and experimental mastery. It allows teachers to take advantage of the benefits of both methods, exploiting the flexibility and safety of simulations while preserving the educational richness of real-world experiments.

Ultimately, analysis of the results shows that the future of science education lies in striking a balance between tradition and innovation. Real experiments remain invaluable for developing students' practical skills, methodological rigor, and critical thinking. Digital simulations, on the other hand, enrich teaching by offering new possibilities, facilitating the visualization of complex phenomena, and overcoming material and organizational constraints. Far from being opposed, these two approaches should be viewed as complementary, in a spirit of educational synergy. The key lies in the ability of teachers to articulate these methods in a coherent manner, according to learning objectives, available resources, and institutional contexts.

Thus, the obvious conclusion is that of a hybrid pedagogy, capable of combining the strength of real experiences with the effectiveness of digital simulations. This integrated approach is not only a response to current constraints, but also a way forward for science education. It enables students to become both practically competent and conceptually sound, preparing them to face the scientific challenges of tomorrow. Teachers play a central role in this process, and providing them with ongoing training is essential to ensuring a successful transition. By combining tradition and innovation, science education can offer students a comprehensive, motivating learning experience that is tailored to the demands of the modern world.
	
\end{document}